\documentstyle[12pt,a4]{article}

\title{Lukyanov's Screening Operators for the Deformed Virasoro Algebra}

\author{
Michio Jimbo\thanks{Department of Mathematics, Faculty of Science,
                            Kyoto University, Kyoto 606, Japan.},
Michael Lashkevich\thanks{L.~D.~Landau Institute for Theoretical Physics,
			    Chernogolovka, 142432, Russia.},\\
Tetsuji Miwa\thanks{Research Institute for Mathematical Sciences,
                            Kyoto University, Kyoto 606, Japan.},
and Yaroslav Pugai$^\dagger$
}

\date{{July 23, 1996, hep-th/9607177}}

\begin{document}

\newcommand{\eqref}[1]{(\ref{#1})}

\newcommand{\mod}{~\hbox{mod}~}
\newcommand{\Bbb}{\bf}
\newcommand{\Z}{{\Bbb Z}}
\newcommand{\R}{{\Bbb R}}
\newcommand{\C}{{\Bbb C}}
\newcommand{\F}{{\cal F}}
\renewcommand{\P}{{\cal P}}
\renewcommand{\H}{{\cal H}}
\newcommand{\Q}{{\cal Q}}
\newcommand{\vep}{\varepsilon}
\newcommand{\ve}{\epsilon}
\newcommand{\bve}{\bar\epsilon}
\newcommand{\p}{\hat\pi}
\newcommand{\half}{{\textstyle{1\over2}}}
\newcommand{\Sym}{\mathop{\rm Sym}\nolimits}
\newcommand{\Ker}{\mathop{\rm Ker}\nolimits}
\renewcommand{\Im}{\mathop{\rm Im}\nolimits}
\newcommand{\Arg}{\mathop{\rm Arg}\nolimits}
\def\leftar#1{\mathrel{\mathop{\longleftarrow}^{#1}}}
\def\rightar#1{\mathrel{\mathop{\longrightarrow}^{#1}}}

\newcommand{\qed}{\hfill \fbox{}\medskip}
\newcommand{\proof}{\medskip\noindent{\it Proof.}\quad }

\newtheorem{thm}{Theorem}
\newtheorem{prop}[thm]{Proposition}
\newtheorem{lem}[thm]{Lemma}
\mathsurround=1mm

\maketitle
\begin{abstract}
The BRST property of Lukyanov's screening operators in the bosonic
representation of the deformed Virasoro algebra is proven.
\end{abstract}

{\it 1.} There are two basic approaches to investigation of
integrable two-dimen\-sio\-nal theories. The most general method
is based on the Yang--Baxter equation. Because of this equation the
theory has an infinite-dimensional abelian symmetry algebra
(commuting integrals of motion)
and admits
studying by use of the Bethe Ansatz technique.
Another approach exploits the
existence of infinite-dimensional non-abelian symmetries (Virasoro algebra,
$W$ algebra, Kac--Moody algebra, etc.)~\cite{BPZ}.
The second approach gives a more detailed
description of the theory, including an explicit computation
of correlation functions. However its applicability was restricted
to conformal models of the 2D quantum field theory.

It was recognized in the work \cite{DFJMN} that
the theory of {\it off-critical} two-dimensional
integrable models of statistical mechanics is tightly
connected with the representation theory of infinite-dimensional
quantum algebras and one can study such models
using ideas of conformal field theory.
In particular, the six-vertex model is governed by
the infinite-dimensional
quantum affine algebra $U_q(\widehat{sl}_2)$.  Namely, the
`half' of the space of states
of the six-vertex model in the corner transfer matrix approach
can be identified with the
irreducible representations of $U_q(\widehat{sl}_2)$.
The well-developed representation theory of the latter
makes it possible to diagonalize the transfer matrix
by the vertex operator technique
and to calculate correlation functions of the theory~\cite{collin}.

In the Andrews--Baxter--Forrester restricted solid-on-solid
(RSOS) models~\cite{ABF} in the regime III
a similar role of dynamical symmetry algebra belongs to the
deformed Virasoro algebra (DVA) discovered
recently~\cite{LukPug1,qVir,LukPug2}.
It is expected by construction~\cite{Luk1,Luk2}
that the structure of  representations of DVA is very similar to
that of the Virasoro algebra. However, it
has not yet been proved rigorously.

The present letter aims at clearing up some open points in this problem.
Recall that,  according to  \cite{FeiFuchs},
the structure of irreducible representations of the
Virasoro algebra heavily depends on the arithmetic nature
of its central charge.  The most complicated (and physically interesting)
case corresponds
to the central charge
\begin{equation}
c=1-{6\over r(r-1)}
\label{eqn:ccharge}
\end{equation}
with integer $r\ge 4$. There are infinitely many singular vectors in the
Verma module in this situation.
An explicit
description of the irreducible representations of the Virasoro algebra
with the central charge (\ref{eqn:ccharge}) in terms of the Fock spaces
is given by use of the two-sided Felder resolution~\cite{Fel89}.
We prove two Propositions
concerning existence of the Felder complex for DVA
as it was claimed in~\cite{LukPug2}.
Namely, we check that {\it certain powers} of the
deformed screening operators introduced by Lukyanov
commute with the generating function of the
deformed Virasoro algebra, {\it i.~e.}\
they are intertwining operators and define the injection structure
of the singular vectors for the latter. We then give a
proof that the deformed screening operators
satisfy the BRST property, so that the
original Felder complex can be deformed to
that for  DVA.

{\it 2.} Recall the definition of the
deformed Virasoro algebra~\cite{qVir}. It is generated by
elements $T_n$, $n\in\Z$, such that the series
$T(z)=\sum_{n\in\Z}T_nz^{-n}$ satisfies the defining relations
\begin{eqnarray}
f\biggl({z\over z'}\biggr)T(z')T(z)\hskip-2.5mm&-&\hskip-2.5mm
f\biggl({z'\over z}\biggr)T(z)T(z')
\cr
&&=(x-x^{-1})[r-1]_x[r]_x\left(\delta\biggl({z\over z'x^2}\biggr)-
\delta\biggl({zx^2\over z'}\biggr)\right).
\label{eqn:deform}
\end{eqnarray}
Here $x$ is a real parameter $0<x<1$, and the following notations
are used:
\begin{eqnarray*}
&&
f(z)={1\over1-z}{(x^{2r}z;x^4)_\infty(x^{-2r+2}z;x^4)_\infty
\over(x^{2r+2}z;x^4)_\infty(x^{-2r+4}z;x^4)_\infty},
\qquad (z;p)_\infty=\prod_{n=0}^\infty (1-zp^n),
\\
&&
\delta(z)=\sum_{n=-\infty}^\infty z^n,
\qquad
[n]_x=\frac{x^n-x^{-n}}{x-x^{-1}}.
\end{eqnarray*}
In the limit $x\to1$ the algebra (\ref{eqn:deform}) gives
the Virasoro algebra with the central charge (\ref{eqn:ccharge}).

The bosonic representation of DVA is formulated in terms of
the Heisenberg algebra generators $\beta_m$, $m\in\Z\backslash\{0\}$,
and the
zero mode operators $\P$ and $\Q$ with the relations
\[
[\beta_m,\beta_{m'}]
=m{[m]_x\over[2m]_x}{[(r-1)m]_x\over[rm]_x}\delta_{m+m',0},
\qquad
[\P,\Q]=-i.
\]
Let us define also the Fock modules
$\F_{l,k}$ generated by $\beta_{-m}$, $m>0$ from
the highest weight vector $|l,k\rangle$:
\[
\P|l,k\rangle=\left(
\sqrt{\textstyle{r\over2(r-1)}}\,l -
\sqrt{\textstyle{r-1\over2r}}\,k
\right)|l,k\rangle,\qquad  \beta_m|l,k\rangle =0,\quad m>0.
\]
For notational convenience we shall use also the operator
$\p=\sqrt{2r(r-1)}\P$.

The generating function $T(z)$ of the deformed Virasoro algebra is given by
\begin{eqnarray}
T(z)&=&\Lambda_+(x^{-1}z)+\Lambda_-(xz)=\sum_{n\in\Z}T_nz^{-n},
\label{eqn:la}
\\
\Lambda_\pm(z)&=&x^{\pm\p }
\mathopen:{\rm e}^{\mp\sum_{m\neq 0}\frac{1}{m}(x^{rm}-x^{-rm})
\beta_m z^{-m}}\mathclose:.
\end{eqnarray}
The colons hereafter mean the Wick ordering both in the
oscillator and zero modes.

Consider now Lukyanov's screening operator~\cite{Luk1,LukPug2}
\begin{eqnarray}
X&=&\oint_{|z|=1}{dz\over2\pi iz}\xi(v){[v+\half-\p]
\over[v-\half]},
\label{eqn:scre}
\\
\xi(v)&=&z^{{r-1\over r}}\mathopen:
{\rm e}^{i\sqrt{{2(r-1)\over r}}(\Q-i\P\log z)
+\sum_{m\neq 0}{\beta_m\over m}(x^m+x^{-m})z^{-m}}\mathclose:,
\end{eqnarray}
where we set $z=x^{2v}$. The notation
$[v]$ stands for the theta function
\[
[v]=x^{v^2/r-v}(x^{2v};x^{2r})_\infty(x^{2(r-v)};x^{2r})_\infty
(x^{2r};x^{2r})_\infty,
\]
having the real half period $r$
\[
[v]=-[v+r].
\]
Note that the integrand of the above expression for $X$
contains the factor~\cite{LukPug2}
$F(v;\p)=[v+\half-\p]/[v-\half]$ which has no counterpart in
the representation theory of the Virasoro algebra.
This factor ensures the integrand to
be single-valued in $z$ and the integration is taken over a closed contour.
In the conformal limit $x\to1$ with $z=x^{2v}$ fixed,
the function $F(v;\p)$ tends to a
constant as a function of $z$ for $\Arg z\neq0$
while the poles and zeros of this function
condense to a cut along the positive real axis in the $z$-plane.

{\it 3.} It was shown in \cite{qVir} that the screening current
\eqref{eqn:scre}
commutes with the deformed Virasoro algebra generators up to a total difference:
\begin{lem}
\label{lem:sh}
\begin{eqnarray}
[T_n,\xi(v)]&=&(x^{r-1}-x^{-r+1})\left(A_n\left(v+\frac{r}{2}\right)
-A_n\left(v-\frac{r}{2}\right)\right),
\\
A_n(v)&=&z^{n+{r-1\over r}}\mathopen:
{\rm e}^{i\sqrt{\frac{2(r-1)}{r}}(\Q-i\P\log z)
+\sum_{m\neq 0}\frac{1}{m}(x^{(r-1)m}
+x^{(-r+1)m})\beta_m z^{-m}}\mathclose:.
\end{eqnarray}
\end{lem}
If the function $F(v;\p)$ did not have poles, Lemma \ref{lem:sh} would
imply the commutativity of $T_n$ and $X$.%
\footnote{In particular, this happens when the screening operator
\eqref{eqn:scre} and the
generators of DVA \eqref{eqn:la} act on the Fock space
$\F_{l,1}$ where the additional factor
$F(v;\p)$  becomes simply $\pm 1$~\cite{qVir}.}
In the general case one needs to be more careful with
the definition of spaces where the screening operators act.
We claim that the following Proposition holds:

\begin{prop}\label{prop:a4.2}
\hfill\break
For $1\leq k\leq r$, the operator $X^k$ commutes with generators
$T_n$
on the space $\F_{l,k'}$ provided $k'\equiv k\pmod r$
\begin{equation}
[T_n,X^k]\vert_{\F_{l,k'}}=0,
\qquad
k'\equiv k \pmod r.
\label{eqn:commute}
\end{equation}
\end{prop}
This Proposition means that in the bosonic realization
the operators $X^k$ acting on the Fock space
$\F_{l,k'}$ are intertwining operators for the deformed Virasoro algebra.

It is important  that the Lukyanov's screening operators
satisfy also the
following property
\begin{prop}\label{prop:a4.1}
\hfill\break
The operator $X$ is nilpotent:
\begin{equation}
X^r=0.
\label{eqn:a4.1}
\end{equation}
\end{prop}
Due to this Proposition one can regard  the
intertwining operators given by powers of the screening operators
as differentials of the two-sided Felder complex
and the Eq. \eqref{eqn:a4.1} can be treated as the BRST property of screening operators.

For the proof of these two Propositions we prepare some lemmas.
In general, consider an expression of the form
\[
\Xi(F)=
\oint\!\cdots\!\oint\frac{dz_1}{2\pi iz_1}\cdots\frac{dz_k}{2\pi iz_k}
\xi(v_1)\cdots \xi(v_k) F(v_1,\cdots,v_k).
\]
Symmetrizing the integrand
and using the commutation relation
\[
\xi(v)\xi(v')=h(v-v')\xi(v')\xi(v),
\qquad h(v)=\frac{[v-1]}{[v+1]},
\]
we can rewrite $\Xi(F)$ into the form $\Xi(\Sym F)$, where
\[
\Sym F(v_1,\cdots,v_k)
=
\frac{1}{k!}\sum_{\sigma\in S_k}
F(v_{\sigma(1)},\cdots,v_{\sigma(k)})
\prod_{i<j\atop\sigma(i)>\sigma(j)}
h(v_{\sigma(i)}-v_{\sigma(j)}).
\]
Set $\hat{F}(v_1,\cdots,v_k)=\Sym F(v_1,\cdots,v_k)$.
In view of the properties $h(v)h(-v)=1$ and $h(0)=-1$,
it is easy to see that
\begin{enumerate}
\renewcommand{\labelenumi}{(\alph{enumi})}
\item
$\hat{F}(v_1,\cdots,v_{i+1},v_i,\cdots,v_k)
=\hat{F}(v_1,\cdots,v_i,v_{i+1},\cdots,v_k)h(v_i-v_{i+1})$,
\item $\hat{F}(v_1,\cdots,v_k)$ has a zero on $v_i=v_j$ ($i<j$).%
\footnote{We assume in (b) that $F(v_1,\cdots,v_k)$ has no poles on $v_i=v_j$.}
\end{enumerate}

The proof of the Propositions will be based on the following identity of
theta functions.
\begin{lem}\label{lem:a4.2}
\begin{equation}
\Sym\prod_{i=1}^k [v_i-2i+2]
=\frac{[k]!}{k![1]^k}
\prod_{i<j}\frac{[v_i-v_j]}{[v_i-v_j-1]}\prod_{i=1}^k[v_i-k+1].
\label{eqn:a4.4}
\end{equation}
Here $[k]!=\prod_{i=1}^k[i]$.
\end{lem}

\proof
Denote by $\hat{F}(v_1,\cdots,v_k)$ the left hand side.
Explicitly this function reads
\begin{equation}
\hat{F}(v_1,\cdots,v_k)=
\frac{1}{k!}\sum_{\sigma\in S_k}
\prod_{i=1}^k [v_{\sigma(i)}-2i+2]
\prod_{i<j\atop\sigma(i)>\sigma(j)}
\frac{[v_{\sigma(i)}-v_{\sigma(j)}-1]}{[v_{\sigma(i)}-v_{\sigma(j)}+1]}.
\label{eqn:a4.2}
\end{equation}
{}In the right hand side of \eqref{eqn:a4.2},
the summand has the same quasi-periodicity property
in each variable $v_i$.
This can be shown by noting that, for any $i\leq 1\leq k$ and $\sigma\in S_k$, we have
\[
\sharp\{ j\mid j>i,~\sigma(j)<\sigma(i)\}
-\sharp\{ j\mid j<i,~\sigma(j)>\sigma(i)\}
=
\sigma(i)-i.
\]
\par
{}From the remark made above, \eqref{eqn:a4.2} has zeroes on $v_i=v_j$ ($i<j$).
Therefore it can be written as
\[
\hat{F}(v_1,\cdots,v_k)=
\prod_{i<j}\frac{[v_i-v_j]}{[v_i-v_j-1]}
G(v_1,\cdots,v_k)
\]
with some holomorphic function $G(v_1,\cdots,v_k)$.

Comparing the quasi-periodicity, we conclude that
\[
G(v_1,\cdots,v_k)=C\prod_{i=1}^k[v_i-k+1]
\]
with some constant $C$.
The constant can be determined by setting $v_i=i+k-1$
to be
\begin{equation}
C=\frac{[k]!}{k![1]^k}.
\label{eqn:consta}
\end{equation}
\qed

\begin{lem}\label{lem:a4.3}
\begin{eqnarray}
X^k&\!=\!&\frac{[k]!}{k![1]^k}
\oint\!\cdots\!\oint\frac{dz_1}{2\pi i z_1}\cdots\frac{dz_k}{2\pi i z_k}
\xi(v_1)\cdots\xi(v_k)
\nonumber
\cr
&&\times\prod_{i<j}\frac{[v_i-v_j]}{[v_i-v_j-1]}
\prod_{i=1}^k\frac{[v_i-\frac{1}{2}+k-\p]}{[v_i-\frac{1}{2}]}.
\label{eqn:miwa}
\end{eqnarray}
\end{lem}

\proof Noting that $\p\xi(v)=\xi(v)(\p-2+2r)$, we see that
\[
X^k=
\oint\!\cdots\!\oint\frac{dz_1}{2\pi i z_1}\cdots\frac{dz_k}{2\pi i z_k}
\xi(v_1)\cdots\xi(v_k)
\prod_{i=1}^k\frac{[v_i+\frac{1}{2}-\p+2k-2i]}{[v_i-\frac{1}{2}]}.
\]
The assertion follows by applying Lemma \ref{lem:a4.2}.
\qed

\noindent {\it Proof of Proposition \ref{prop:a4.2}.}\quad
Note first that
$[v_i-\frac{1}{2}+k-\p] =\pm [v_i-\frac{1}{2}]$ on
${\cal F}_{l,k'}$, provided $k'\equiv k\pmod r$.
Under this circumstance, we get from Lemmas \ref{lem:sh} and
\ref{lem:a4.3} that
\begin{eqnarray}
&&[T_n,X^k]\vert_{{\cal F}_{l,k'}}
\nonumber\\
&&\propto
\sum_{s=1}^k
\oint\!\cdots\!\oint\frac{dz_1}{2\pi i z_1}\cdots\frac{dz_k}{2\pi i z_k}
\xi(v_1)\cdots\left(A_n\left(v_s+\frac{r}{2}\right)
-A_n\left(v_s-\frac{r}{2}\right)\right)\cdots
\xi(v_k)
\nonumber \\
&&
\times\prod_{i<j}\frac{[v_i-v_j]}{[v_i-v_j-1]}.
\nonumber\\
&&\label{eqn:a4.5}
\end{eqnarray}
The integral is taken over the contours $|z_1|=\cdots=|z_k|=1$.
With the change of variable $z_s\rightarrow x^{-2r}z_s$,
each term in the right hand side of \eqref{eqn:a4.5}
formally vanishes.
It remains to show that the poles $[v_i-v_j-1]=0$ do not affect the
procedure.

To see this, note the following normal ordering rules
\begin{eqnarray*}
\xi(v_1)A_n(v_2)&=&z_1^{\frac{2(r-1)}{r}}
\frac{(x^{-r+2}z_2/z_1;x^{2r})_\infty}{(x^{3r-2}z_2/z_1;x^{2r})_\infty}
\mathopen:\xi(v_1)A_n(v_2)\mathclose:,
\\
A_n(v_2)\xi(v_1)&=&
z_2^{2\frac{(r-1)}{r}}
\frac{(x^{-r+2}z_1/z_2;x^{2r})_\infty}{(x^{3r-2}z_1/z_2;x^{2r})_\infty}
\mathopen:\xi(v_1)A_n(v_2)\mathclose:.
\end{eqnarray*}
Using these equations we find that the expressions
\[
\frac{[v_i-v_s]}{[v_i-v_s-1]}\xi(v_i)A_n\left(v_s+\frac{r}{2}\right)
~(i<s),
\quad
\frac{[v_s-v_i]}{[v_s-v_i-1]}A_n\left(v_s+\frac{r}{2}\right)\xi(v_i)
~(i>s)
\]
have poles only at
$z_s=x^{2rj-2}z_i$ ($j=1,2,\cdots$) and
$z_s=x^{-2rj+2}z_i$ ($j=2,3,\cdots$).
Therefore, the contour for $z_s$ in the
\eqref{eqn:a4.5} can be shifted from $|z_s|=1$ to $|z_s|=x^{-2r}$ without
affecting the integral.
This completes the proof of the Proposition~\ref{prop:a4.2}.
\qed

\noindent {\it Proof of Proposition \ref{prop:a4.1}.}\quad
This is an immediate consequence of Lemma \ref{lem:a4.3} .
Indeed, taking $k=r$ one obtains that the constant factor
in \eqref{eqn:miwa} vanishes since $[r]=0$.
\qed

\noindent{\it Remark 1.}\quad
Because of the $r\rightarrow 1-r$ symmetry in the DVA defining
relations \eqref{eqn:deform}
there is the second screening operator
defined by
\[
\tilde{X}=\oint_{|z|=1}\frac{dz}{2\pi i z}
\tilde{\xi}(v)\frac{[v-\frac{1}{2}+\p]'}{[v+\frac{1}{2}]'}
\]
where
\begin{eqnarray*}
&&\tilde{\xi}(v)=
{\rm e}^{-i\sqrt{\frac{2r}{r-1}}(\Q-i\P\log z)}
\mathopen:{\rm e}^{-\sum_{m\neq 0}\frac{1}{m}(x^m+x^{-m})
\tilde{\beta}_mz^{-m}}\mathclose:,
\\
&&\tilde{\beta}_m=\frac{[rm]_x}{[(r-1)m]_x}\beta_m,
\end{eqnarray*}
and
\[
[v]'=x^{\frac{v^2}{r-1}-v}(x^{2v};x^{2r-2})_\infty
(x^{2(r-v-1)};x^{2r-2})_\infty(x^{2r-2};x^{2r-2})_\infty \quad .
\]
It can be shown that
\begin{enumerate}
\renewcommand{\labelenumi}{(\alph{enumi})}
\item $[X,\tilde{X}]=0$,
\item $\tilde{X}^{r-1}=0$,
\item $[T_n,\tilde{X}^l]=0$ on ${\cal F}_{l',k}$
provided $l'\equiv l\pmod {r-1}$.
\end{enumerate}
We omit the proofs of these statements since they are completely analogous to those
provided above.

\noindent{\it Remark 2.}\quad The proof can be easily generalized to
arbitrary real values of $r>1$.
In the same way as above,
the operator $X^k$ commutes with $T_n$ on the space $\F_{l,k}$
(where $k$ is a positive integer)
$$
[T_n,X^k]|_{\F_{l,k}}=0.
\eqno(\ref{eqn:commute}')
$$
For $r$ irrational, this property is sufficient to construct the
BRST complex.
For rational $r$,
\begin{equation}\label{eqn:ratr}
r=\frac{q}{q-p}
\end{equation}
with coprime positive integers $p$ and $q$, $q>p$,
the main propositions take the form
\medskip

\noindent{\bf Proposition $\bf 2'$}{\it
\hfill\break
{\mathsurround=0pt
The operator $X^k$ commutes with $T_n$
on the space $\F_{l,k'}$ provided $k'\equiv k\pmod q$:}
$$
[T_n,X^k]\vert_{\F_{l,k'}}=0,
\qquad
k'\equiv k \pmod q.
\eqno(\ref{eqn:commute}'')
$$
}

\noindent{\bf Proposition $\bf 3'$}{\it
\hfill\break
The operator $X$ is nilpotent:
$$
X^q=0.
\eqno(\ref{eqn:a4.1}')
$$
}

Similar properties hold for $\tilde X$, but $\tilde X^p=0$ provided $p>1$.

{\it 4.} We have proved that Lukyanov's screening operators
satisfy the BRST property. Note that, in the $x=1$ case,
the symmetrization of the integrand in $X^k$ also leads to an
appearance of a constant factor%
\footnote{Compare with \eqref{eqn:consta}.}
\[
\frac{1}{k!}\prod_{j=1}^k\frac{e^{2\pi i j \frac{r-1}{r}}-1}{e^{2\pi i  \frac{r-1}{r}}-1}
\]
which becomes zero at $k=r$
and the $r$th power of the screening operator vanishes~\cite{Fel89}.
As it was demonstrated above, the situation
in the deformed case turns out to be more complicated
and the property \eqref{eqn:a4.1} of Lukyanov's screening operators follows
as a result of non-trivial theta function identities.

On the basis of Propositions 2 and 3,
we can construct a family of BRST complexes of DVA
depending on the continuous parameter $x$:
\[
\cdots\rightar{X^k}\F_{l,2r-k}
\rightar{X^{r-k}}\F_{l,k}
\rightar{X^{k}}\F_{l,-k}
\rightar{X^{r-k}}\cdots.
\]
In the undeformed case $x=1$, it is known that the
latter complex has trivial
cohomologies except at the term $\F_{l,k}$ while
$\Ker_{\F_{l,k}}{X^k}/\Im_{\F_{l,k}}{X^{r-k}}$
provides an irreducible representation of the Virasoro algebra~\cite{Fel89}.
Note that the derivation of this result relies entirely on the
structure theory of representations of the Virasoro algebra, and
no direct proof is available.
In the deformed case, we expect that the same is true about the
cohomology, at least for generic values of $x$.
However, as very little is known about the representations of DVA,
a rigorous proof is still lacking.

Among the open problems we would like to
mention the following one. Recall that
the braiding matrices of the chiral vertex operators
of the Virasoro algebra are
constant solutions of the Yang--Baxter equation in the IRF form~\cite{DotsFat}.
This reflects the existence of a hidden quantum group symmetry
in the minimal models of conformal field theory.%
\footnote{Here we restrict our attention to one chirality of the
conformal field theory.} More explicitly,
in the bosonic realization, the screening operators
can be regarded as giving a representation of nilpotent generators
of two quantum groups $U_q(sl_2)$
with the deformation parameters related by the $r\rightarrow  1-r$
transformation~\cite{DotsFat}. The action of the
Virasoro algebra generators preserves the monodromy of matrix elements
of vertex operators,
and the complete symmetry algebra
is a tensor product of the infinite-dimensional part (the Virasoro algebra)
and the finite-dimensional one given by two quantum groups.
The monodromy of matrix elements of vertex operators in the ABF models
is given by an elliptic solution of the Yang--Baxter equation~\cite{Fodal94}.
Owing to the works \cite{LukPug2,qVir}
it becomes clear that the proper deformation
of Virasoro algebra is given by DVA.
However, it is not clear yet what is the ``elliptic deformation"
of the quantum group part of the symmetry algebra
of critical theory.

{}From this point of view our results on the properties of deformed
screening operators can be considered as a first step
toward understanding this problem.

\section*{Acknowledgments}

M.~J. and T.~M.  are grateful to L.~D.~Landau
Institute for Theoretical Physics and especially to A. Belavin for kind
hospitality.
Ya.~P.\ and M.~L.\ are pleased to acknowledge fruitful
discussions with  A.~Belavin, B.~Feigin, A.~Kadeishvili, and S.~Lukyanov.
Ya.~P.\ would like to thank Research Institute for Mathematical Sciences
for kind hospitality. M.~L.\ was supported in part by INTAS under the grant
CNRS~1010-CT93-0023.


\end{document}